\documentstyle[aps,epsfig,amssymb,floats]{revtex}

\begin{document}
\draft

\twocolumn[\hsize\textwidth\columnwidth\hsize\csname
@twocolumnfalse\endcsname

\title{First-principles study of structural, vibrational and
lattice dielectric properties of hafnium oxide} 
\author{Xinyuan Zhao and David Vanderbilt}
\address{Department of Physics and Astronomy, Rutgers University,
	Piscataway, NJ 08854-8019}
\date{February 25, 2002}
\maketitle
\begin{abstract}
Crystalline structures, zone-center phonon modes, and the related
dielectric response of the three low-pressure phases of HfO$_2$ have
been investigated in density-functional 
theory using ultrasoft pseudopotentials and a plane-wave basis.
The structures of low-pressure HfO$_2$ polymorphs are carefully studied
with both the local-density approximation (LDA) and the
generalized gradient approximation (GGA).  The
fully relaxed structures obtained with either exchange-correlation 
scheme agree reasonably well with experiment, although LDA yields 
better overall agreement. After calculating the Born effective charge
tensors and the force-constant matrices by finite-difference
methods, the lattice dielectric susceptibility tensors for
the three HfO$_2$ phases are computed by decomposing the tensors into the
contributions from individual infrared-active phonon modes. 
\end{abstract}


\vskip2pc]

\columnseprule 0pt
\narrowtext

\begin{figure}[b!]
\begin{center}
   \epsfig{file=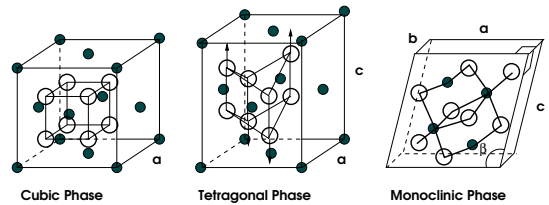, width=2.8in}
\end{center}
\caption{Structures of the three HfO$_2$ phases. Small dark circles
and larger open circles denote Hf and O atoms respectively.  Hf--O
bonds are only shown in $m$-HfO$_2$. In $t$-HfO$_2$, the arrows
indicate the shift of oxygen pairs.}
\label{fig:struct} 
\end{figure}

Hafnia (HfO$_2$) is technologically important because of its
extraordinary high bulk modulus, high melting point, and high chemical
stability, as well as its high neutron absorption cross
section. HfO$_2$ 
resembles its twin oxide, zirconia (ZrO$_2$), in many
physical and chemical properties. The resemblance is attributable
to the structural similarity between the two oxides, which can
in turn be explained by the chemical similarity of Hf and Zr, which have
similar atomic and ionic radii (i.e., ionic radii for Hf$^{4+}$ and Zr$^{4+}$
of 0.78 and 0.79\,${\rm \AA}$, respectively \cite{ruh70}) as a result of
the so-called lanthanide contraction.  Under ambient pressure, both
oxides are monoclinic ($m$, space group $P2_{1}/c$) at low temperature,  
and transform to a tetragonal structure ($t$, space group $P4_2/nmc$)
and then to a cubic structure ($c$, space group $Fm3m$) as the
temperature increases, as illustrated in Fig.~\ref{fig:struct}.

High-K metal-oxide dielectrics have recently been the focus of
substantial ongoing efforts directed toward finding a replacement
for SiO$_2$ as the gate dielectric in complementary
metal-oxide-semiconductor (CMOS) devices. HfO$_2$, ZrO$_2$ and
their SiO$_2$ mixtures show promise for this purpose
\cite{wilk,gusev}. Thus, a systematic theoretical investigation of
the structural and dielectric properties of these dielectrics, in
both bulk and thin-film form, is clearly desirable. As a first step
in this direction, we have, in a previous paper \cite{zhao},
investigated the bulk structures and lattice dielectric response of
ZrO$_2$ polymorphs. We found that the dielectric
responses vary dramatically with the crystal phase.  Specifically,
we found that the monoclinic phase has a strongly anisotropic
lattice dielectric tensor and a rather small
orientationally-averaged dielectric constant owing to the fact that
the mode effective charges associated with the softest modes are
relatively weak.

This report presents the corresponding work on HfO$_2$, providing the
first thorough theoretical study of the structural, vibrational and
lattice dielectric properties of the HfO$_2$ phases. Such
properties are naturally expected to be similar to those of ZrO$_2$
in view of the chemical similarities mentioned above.  We find that
this is generally true, although we also find some significant
quantitative differences in some of the calculated properties.

The calculation of the lattice contributions to the static dielectric
tensor $\epsilon_{0}$ entails the computations of the Born effective
charge tensors ${\bf Z}^*$ and the force-constant matrices
${\bf \Phi}$.  The ${\bf Z}^*$ tensors, defined via
\hbox{$\Delta {\bf P} = (e/V)\, \sum_i {\bf Z}_{i}^{*} \cdot
\Delta {\bf u}_{i}$}
, are obtained by finite 
differences of polarizations ({\bf P}) as various sublattice
displacements (${\bf u}_{i}$) are imposed, with the electronic part of
the polarizations computed using the Berry-phase approach
\cite{modern-pol,born-charge}. Here $V$ is the volume of the unit
cell, $e$ is the electron charge, and $i$ labels the atom in the unit
cell. We then calculate the force-constant matrix,
\hbox{$\Phi^{\alpha \beta}_{ij} = - \partial F^{\alpha}_{i} / \partial
u^{\beta}_{j} \simeq - \Delta F^{\alpha}_{i} / \Delta
u^{\beta}_{j}$}
by calculating all the Hellmann-Feynman forces
$F^{\alpha}_{i}$ caused by making displacements $u^{\beta}_{j}$
of each atom in each Cartesian direction in turn
(Greek indices label the Cartesian coordinates).
The resulting $\Phi$ matrix is symmetrized to clean up
numerical errors, the dynamical matrix 
$D^{\alpha \beta}_{ij}=(M_iM_j)^{-1/2}\,\Phi^{\alpha \beta}_{ij}$
is constructed, and the latter is then diagonalized to obtain the
eigenvalues $\omega_\lambda^2$ and eigenvectors $\xi_{i,\lambda \beta}$.

The static dielectric tensor can be decomposed into a contribution
$\epsilon_{\infty}$ arising from purely electronic screening, and the
contributions of the IR-active phonon modes, according to \cite{explan-strain}
\begin{equation}
\epsilon^{0}_{\alpha \beta} = \epsilon^{\infty}_{\alpha
\beta} + \frac{4 \pi e^{2}}{M_0\,V} \sum_{\lambda} \frac{
 {\widetilde{Z}}^{*}_{\lambda \alpha} \, {\widetilde{Z}}^{*}_{\lambda \beta}
}{\omega_{\lambda}^{2}} \;.
\label{eq:lattcont}
\end{equation}
Here the
$\widetilde{Z}^{*}_{\lambda \alpha} = \sum_{i \beta} \, 
Z^{*}_{i, \alpha \beta} \, \left({M_0}/{M_i}\right)^{1/2} \, 
\xi_{i,\lambda \beta}$ are mode effective charges, $e$ is the
electron charge, $M_0$ is a reference mass that we take for
convenience to be 1\,amu, $\omega_{\lambda}$ is the frequency of the
$\lambda$-th IR-active phonon mode, and $V$ is the volume of
the 3-atom, 6-atom, or 12-atom unit cell for cubic, tetragonal,
or monoclinic cases, respectively. $\xi_{i,\lambda \beta}$, the
eigendisplacement of atom $i$ in phonon mode $\lambda$, is normalized
according to $\sum_{i \alpha} \xi_{i,\lambda \alpha} \,
\xi_{i,\lambda^{'} \alpha} = \delta_{\lambda \lambda^{'}}$.

\begin{table}
\caption{Calculated structural parameters for three HfO$_2$ phases
using both LDA and GGA. Lattice 
parameters $a$, $b$, $c$ are in ${\rm \AA}$, $\beta$ is in degrees,
and $V$ (volume per formula) is in ${\rm \AA}^{3}$. Internal 
coordinates $x$, $y$ and $z$ are dimensionless.} 
\begin{tabular}{cddddd}
& Present & Present & Previous & & ZrO$_2$ \\
& LDA & GGA & LDA\tablenotemark[1] &
   \multicolumn{1}{c}{Expt.\tablenotemark[2]} & LDA\tablenotemark[3] \\
\tableline
\multicolumn{6}{l}{Cubic} \\
$V$     & 31.95  & 36.15  & 32.01 & 32.77 & 31.95 \\
 a      & 5.037  & 5.248  & 5.04 & 5.08  & 5.037 \\
\tableline
\multicolumn{6}{l}{Tetragonal} \\
$V$     & 32.77  & 37.74  & 32.5 &   & 32.26 \\
 a      & 5.056  & 5.299  & 5.03 &   & 5.029 \\
 c      & 5.127  & 5.373  & 5.11 &   & 5.100 \\
$d_{z}$ & 0.042  & 0.041  & 0.038 &   & 0.041  \\
\tableline
\multicolumn{6}{l}{Monoclinic} \\
$V$             & 34.35 & 38.01 & 33.9 & 34.58 & 34.35\\
a               & 5.106 & 5.291 & 5.08 & 5.117 & 5.108\\
b               & 5.165 & 5.405 & 5.19 & 5.175 & 5.170\\
c               & 5.281 & 5.366 & 5.22 & 5.291 & 5.272\\
$\beta$         & 99.35 & 97.92 & 99.77 & 99.22 & 99.21\\
$x_{\rm Hf}$    & 0.280 & 0.276 & 0.280 & 0.276 & 0.277\\
$y_{\rm Hf}$    & 0.043 & 0.039 & 0.044 & 0.040 & 0.042\\
$z_{\rm Hf}$    & 0.209 & 0.209 & 0.208 & 0.208 & 0.210\\
$x_{\rm O_{1}}$ & 0.076 & 0.089 & 0.078 & 0.074 & 0.069\\
$y_{\rm O_{1}}$ & 0.346 & 0.367 & 0.350 & 0.332 & 0.333\\
$z_{\rm O_{1}}$ & 0.337 & 0.317 & 0.332 & 0.347 & 0.345\\
$x_{\rm O_{2}}$ & 0.447 & 0.447 & 0.446 & 0.449 & 0.450\\
$y_{\rm O_{2}}$ & 0.759 & 0.762 & 0.759 & 0.758 & 0.757\\
$z_{\rm O_{2}}$ & 0.483 & 0.488 & 0.485 & 0.480 & 0.480
\end{tabular}
\tablenotetext[1] {Ref.~\onlinecite{demkov}.}
\tablenotetext[2] {Ref.~\onlinecite{cubic} for cubic;
Ref.~\onlinecite{mono} for monoclinic.}
\tablenotetext[3] {Ref.~\onlinecite{zhao}.}
\label{table:parameters}
\end{table}

The calculations are carried out within an ultrasoft pseudopotential
\cite{uspp} implementation of density-functional 
theory with a plane-wave basis
and a conjugate-gradient minimization algorithm. The crystal
structures of HfO$_2$ polymorphs are investigated in the
local-density approximation (LDA) as parameterized by Ceperley and
Alder \cite{lda-ca} as well as in the generalized gradient
approximation (GGA) using PBE parametrization \cite{gga-pbe}.  We
find that LDA yields slightly better agreement with the
experimental structures, and we therefore suggest that our LDA
results for the dielectric properties are more reliable.
The $4s$ and $4p$ semicore shells are
included in the valence in the Hf pseudopotential, and an energy
cutoff of 25\,Ry is chosen.  A 4$\times$4$\times$4 Monkhorst-Pack
k-point mesh is found to provide sufficient precision in the
calculations of total energies and forces, and a
4$\times$4$\times$20 k-point sampling is used for calculating the
Berry-phase polarization \cite{modern-pol}.  Each atomic sublattice
is displaced in turn along each Cartesian direction by $\pm$0.2\% in
lattice units, the electronic polarization and Hellmann-Feynman
forces are computed, and ${\bf Z}^*$ and ${\bf \Phi}$,
are then constructed by finite differences from the results.

Tabulated in Table~\ref{table:parameters} are the relaxed
structural parameters for the three HfO$_2$ polymorphs, with the
corresponding data for ZrO$_2$ listed in the last column for
comparison \cite{zhao}. While several structural determinations for
$m$-HfO$_2$ can be found in the literature \cite{ruh70,mono},
corresponding results for the tetragonal and cubic phases are
relatively sparse \cite{cubic}.  Nor has there been much
theoretical work on hafnia; most important is the recent work
of Ref.~\cite{demkov} which agrees quite well with our results. For
$m$-HfO$_2$, the parameters given in Ref.~\cite{mono} were used as the
starting point of our relaxation procedures, while for $t$- and
$c$-HfO$_2$ we started the relaxation from the zirconia experimental
structures. It can readily be seen that both the LDA and GGA agree
reasonably well with the previous work, but that the LDA yields a better
overall agreement. Our total-energy calculations reproduce the correct
energetic ordering of the phases (monoclinic then tetragonal then cubic)
using either LDA or GGA.

\begin{table}
\caption{LDA dynamical effective charges ${\bf Z}^*$ for HfO$_2$ phases.
(Values in parentheses are GGA results.)}
\begin{tabular}{cdddddd}
&\multicolumn{2}{c}{Hf} &\multicolumn{2}{c}{O$_1$} &\multicolumn{2}{c}{O$_2$} \\
\tableline
Cubic &\multicolumn{2}{c}{5.85} &\multicolumn{2}{c}{-2.93}
      &\multicolumn{2}{c}{-2.93}  \\ 
\tableline
\multicolumn{7}{l}{Tetragonal} \\
$x'x'$ & \multicolumn{2}{c}{5.84} & \multicolumn{2}{c}{-3.53}
       & \multicolumn{2}{c}{-2.31} \\ 
$y'y'$ & \multicolumn{2}{c}{5.84} & \multicolumn{2}{c}{-2.31}
       & \multicolumn{2}{c}{-3.53} \\ 
$zz$   & \multicolumn{2}{c}{5.00} & \multicolumn{2}{c}{-2.50}
       & \multicolumn{2}{c}{-2.50} \\  
\tableline 
\multicolumn{7}{l}{Monoclinic} \\
      $xx$ &  5.56 & ( 5.57) & -3.09 & (-3.10) & -2.48 & (-2.47) \\
      $xy$ & -0.47 & (-0.56) &  0.97 & ( 0.90) &  0.20 & ( 0.15) \\
      $xz$ &  0.96 & ( 0.91) & -0.58 & (-0.53) & -0.39 & (-0.36) \\
      $yx$ & -0.13 & (-0.02) &  1.37 & ( 1.29) &  0.21 & ( 0.11) \\
      $yy$ &  5.55 & ( 5.57) & -2.73 & (-2.70) & -2.82 & (-2.87) \\
      $yz$ &  0.14 & ( 0.07) & -0.71 & (-0.61) &  0.35 & ( 0.40) \\
      $zx$ &  0.21 & ( 0.27) & -0.18 & (-0.20) & -0.07 & (-0.09) \\
      $zy$ &  0.41 & ( 0.45) & -0.61 & (-0.51) &  0.43 & ( 0.46) \\
      $zz$ &  4.74 & ( 4.64) & -2.24 & (-2.16) & -2.58 & (-2.52) \\
\end{tabular}
\label{table:zstar}
\end{table}
Our results for the dynamical effective charges are presented in
Table~\ref{table:zstar}. The symmetry of $c$-HfO$_2$ requires that
${\bf Z}^*$ be isotropic on each atom. In $t$-HfO$_2$, the
shifting of oxygen atoms creates two different configurations for
oxygen atoms (denoted O$_1$ and O$_2$) and introduces off-diagonal
elements. Thus, it is more natural to refer to a reference frame
$x'$-$y'$-$z$ that is rotated 45$^\circ$ about the $\hat{z}$
axis from the original Cartesian frame. ${\bf Z}^*$(O$_{1,2}$) 
become diagonal in this frame.  In $m$-HfO$_2$, there are two 
non-equivalent oxygen sites (i.e., the 3-fold and 4-fold 
oxygens, labeled as O$_1$ and O$_2$ respectively). The crystal
structure can then be regarded as composed of three kinds of atoms,
namely, Zr, O$_{1}$, and O$_{2}$, which all have equally low 
symmetry, and their resulting ${\bf Z}^*$ tensors are neither
diagonal nor symmetric. The presence of two non-equivalent oxygen
atoms with very different environments is reflected in the
difference between the Born effective charge tensors for O$_1$ and
O$_2$. The anomalously large
$Z^{*}$ values indicate that there is a strong dynamic charge transfer
along the Hf$-$O bond as the bond length varies, indicating a mixed
ionic-covalent nature of the Hf$-$O bond. The resultant relatively
delocalized distribution of the electronic charge is  
very similar to ZrO$_2$, and is quite common in partially covalent oxides.

\begin{table}
\caption{Theoretical (LDA and GGA) and experimental
(Ref.~\protect\onlinecite{arashi}) frequencies (in cm\,$^{-1}$) of
Raman-active phonon modes in monoclinic HfO$_{2}$.}
\begin{tabular}{ccccc}
 Irrep & Mode & LDA & GGA & Expt.~\protect\onlinecite{arashi} \\ 
\tableline
$A_{g}$ & 1 & 128 & 125 & 113 \\
        & 2 & 142 & 132 & 133 \\
	& 3 & 152 & 171 & 149 \\
	& 4 & 261 & 248 & 256 \\
        & 5 & 326 & 339 & $\phantom{^a}$323\tablenotemark[1] \\
	& 6 & 423 & 382 & 382 \\
	& 7 & 514 & 440 & 498 \\
	& 8 & 608 & 557 & 577 \\
	& 9 & 738 & 640 & 672 \\
\tableline
$B_{g}$ & 1 & 131 & 120 & 133 \\
 	& 2 & 175 & 152 & 164 \\
	& 3 & 250 & 223 & 242 \\
	& 4 & 380 & 318 & 336 \\
	& 5 & 424 & 385 & 398 \\
	& 6 & 533 & 466 & 520 \\
	& 7 & 570 & 529 & 551 \\
	& 8 & 667 & 627 & 640 \\
	& 9 & 821 & 716 & $\phantom{^a}$773\tablenotemark[2]\\
\end{tabular}
\tablenotetext[1] {Unassigned.}
\tablenotetext[2] {Ref.~\onlinecite{jayaraman}.}
\label{table:raman}
\end{table}

Since HfO$_2$ is isomorphic to ZrO$_2$, the analysis of the phonon
modes at $\Gamma$ is the same for HfO$_2$ as for ZrO$_2$ 
\cite{zhao}. Of 36 phonon modes predicted for $m$-HfO$_2$, 18
modes ($9A_{g} + 9 B_{g}$) are Raman-active and 15 modes ($8 A_{u} + 7
B_{u}$) are IR-active, the remaining three modes being the
zero-frequency translational modes. There are three IR-active modes
($A_{2u}$ and two $E_u$) and three Raman-active modes ($A_{1g}$,
$B_{1g}$ and $E_{g}$) for $t$-HfO$_2$. Only one IR-active mode
(one $T_{1u}$ triplet) is predicted for $c$-HfO$_2$.

The Raman spectra of $m$-HfO$_2$ have been extensively measured
experimentally \cite{asher,arashi,carlone,jayaraman}, but the situation
is not entirely satisfactory \cite{carlone}.  Issues concerning the
number of modes and the mode assignments still remain unresolved,
partially because of sample impurities and the broadness and weakness
of some observed features. Thus, our {\it ab-initio} theoretical
calculation can play an important role in establishing the Raman
assignments. Table~\ref{table:raman} shows the frequencies of the
$A_g$ and $B_g$ Raman-active modes as calculated in LDA and GGA,
together with the observed frequencies from a polarized Raman measurement
on a high-quality single crystal\cite{arashi}.  The agreement is
generally excellent; the observed Raman shifts mostly fall
comfortably in the LDA--GGA range.  A later single-crystal (but
unpolarized) Raman spectrum\cite{jayaraman} shows almost identical
mode frequencies.  However, a few details about the Table deserve comment.
(i) We omit the weak mode reported as $A_g$ at 268\,cm\,$^{-1}$ in
Ref.~\cite{arashi} because it is not confirmed in Ref.~\cite{jayaraman}
and it does not fit with our theoretical assignments.
(ii) We assign the 323\,cm\,$^{-1}$ mode observed in
Refs.~\cite{arashi,jayaraman} as $A_g$.
(iii) The feature observed at 872\,cm$^{-1}$ in Ref.~\cite{arashi} is
presumed to be a two-phonon process and is omitted.
(iv) A weak mode is observed at  773\,cm$^{-1}$ in Ref.~\cite{jayaraman};
since this is consistent with our highest-frequency $B_g$ mode, we
assign it as such.

\begin{table}
\caption{ Frequencies $\omega_\lambda$ (cm$^{-1}$) and
scalar mode effective charges ${\widetilde{Z}}^{*}_{\lambda}$ of
IR-active phonon modes for HfO$_{2}$ phases, where
${\widetilde{Z}}^{*\,2}_{\lambda} = \sum_\alpha 
{\widetilde{Z}}^{*\,2}_{\lambda \alpha}$.}  
\begin{tabular}{cccccc}
 & & \multicolumn{2}{c}{LDA} & \multicolumn{2}{c}{GGA} \\
 & Irrep & $\omega_{\lambda}$ & ${\widetilde{Z}}^{*}_{\lambda}$ &
$\omega_{\lambda}$ & ${\widetilde{Z}}^{*}_{\lambda}$ \\
\tableline
Cubic & & & & &  \\
1 & $T_{1u}$ & 286 & 1.12 & & \\
\tableline
\multicolumn{5}{l}{Tetragonal} \\
1 & $E_{u}$ & 117 & 1.26 & &  \\
2 & $A_{2u}$& 384 & 1.45 & &  \\
3 & $E_{u}$ & 536 & 1.13 & &  \\
\tableline
\multicolumn{5}{l}{Monoclinic} \\
  1  & $A_{u}$ & 140 & 0.049 & 123 & 0.075 \\
  2  & $A_{u}$ & 190 & 0.003 & 162 & 0.063 \\
  3  & $B_{u}$ & 246 & 0.887 & 223 & 0.823 \\
  4  & $A_{u}$ & 255 & 0.764 & 250 & 0.917 \\
  5  & $B_{u}$ & 262 & 0.121 & 252 & 0.297 \\
  6  & $B_{u}$ & 354 & 1.623 & 300 & 1.791 \\
  7  & $B_{u}$ & 378 & 1.126 & 331 & 1.081 \\
  8  & $A_{u}$ & 393 & 1.148 & 360 & 1.196 \\
  9  & $A_{u}$ & 445 & 1.218 & 391 & 1.226 \\
  10 & $B_{u}$ & 449 & 1.497 & 414 & 1.339 \\
  11 & $A_{u}$ & 529 & 0.836 & 456 & 0.676 \\
  12 & $B_{u}$ & 553 & 0.810 & 494 & 0.814 \\
  13 & $A_{u}$ & 661 & 0.788 & 577 & 0.962 \\
  14 & $A_{u}$ & 683 & 0.688 & 634 & 0.032 \\
  15 & $B_{u}$ & 779 & 0.997 & 694 & 0.900 \\
\end{tabular}
\label{table:infrared}
\end{table}

The frequencies of the IR-active phonon modes for the three
HfO$_2$ phases are tabulated in Table~\ref{table:infrared},
together with the scalar mode effective charges. It can be
seen that the frequencies calculated in GGA are shifted
to lower frequency by $\sim$10\,--\,16\% relative to the
LDA ones, while the mode assignments coincide exactly.
As indicated in Eq.~(\ref{eq:lattcont}), the contribution of a
given IR-active mode to the static dielectric constant scales as
$\widetilde{Z}^{*\,2} / \omega_{\lambda}^2$ \cite{zhao}, so that
one or more low-frequency modes with large $\widetilde{Z}^{*}$'s
are needed to yield a large dielectric constant. As can be seen
from Table~\ref{table:infrared}, however, the few softest modes
($<$ 300\,cm$^{-1}$) have relatively small $\widetilde{Z}^{*}$'s,
while the more active infrared modes come in the intermediate
range of the IR spectrum (350\,--\,450\,cm$^{-1}$).  The
general pattern is very similar to the case of ZrO$_2$.

The lattice contributions to the dielectric tensors are obtained
by summing the second term of Eq.~(\ref{eq:lattcont}) over all
the IR-active modes.  Using the LDA we find
\[\epsilon^{\rm latt}_{\rm cubic} =  \left( \begin{array}{ccc}
        23.9 & 0  & 0 \\
        0  & 23.9 & 0 \\
        0  & 0 & 23.9 \\
          \end{array} \right),  \]

\[\epsilon^{\rm latt}_{\rm tetra}  =  \left( \begin{array}{ccc}
        92.3 & 0 & 0\\
        0 & 92.3 & 0 \\
        0 & 0 & 10.7 \\
          \end{array} \right),  \]

\[\epsilon^{\rm latt}_{\rm mono}  =  \left( \begin{array}{ccc}
        13.1 & 0 & 1.82\\
        0 & 10.8 & 0 \\
        1.82 & 0 & 7.53 \\
          \end{array} \right).  \]
(The corresponding matrix elements of $\epsilon^{\rm latt}_{\rm mono}$
in the GGA tend to be larger than the LDA results by $\sim$18\%.)
When compared with ZrO$_2$, the off-diagonal elements of 
$\epsilon^{\rm latt}_{\rm mono}$ are roughly doubled, while the 
diagonal elements become smaller.
Most surprisingly, the $x$-$y$ components of
$\epsilon^{\rm latt}_{\rm tetra}$ become more than twice as large as
for ZrO$_2$, while the $z$ component decreases by $\sim$28\%.
We find the isotropic $\epsilon^{\rm latt}_{\rm cubic}$ to be
23.9, somewhat smaller than the value of 31.8 for ZrO$_2$ \cite{zhao}.  

A direct comparison of these dielectric tensors with experiment
is not feasible since there are few experimental measurements, 
especially on the cubic and tetragonal phases.
On the other hand, most measurements of which we are aware have been
carried out on thin films (presumed to be monoclinic),
and the reported dielectric constants span a wide range of
16\,--\,45 \cite{gusev,harrop,kukli}. Assuming an isotropic
$\epsilon_\infty\simeq5$ \cite{zhao}, we obtained orientationally
averaged static dielectric constants of 29, 70, and 16 (18 in
GGA) for the cubic, tetragonal and monoclinic HfO$_2$ phases,
respectively. Our results then agree reasonably well with the more
recent results in Ref.~\cite{kukli}
(thin film $\sim$ 1700\,${\rm \AA}$) and Ref. \cite{gusev}
(ultrathin film $<$ 100\,${\rm \AA}$) which report $\epsilon_0$
to be 16 and 20 respectively. The surprising high $\epsilon_0$
measured in other experiments could possibly be explained by the
presence of $t$-HfO2, which is known to be a metastable phase
and which might be stablized by film stress or grain-size effects
\cite{cubic,kukli,garvie}.

In summary, we have investigated here the Born effective charge
tensors, zone-centered phonons, and the lattice contributions to the
static dielectric tensors for the three HfO$_2$ phases. It is found that the
cubic and tetragonal phases have much larger dielectric response than
the monoclinic phase, with an even stronger anisotropy in $t$-HfO$_2$
than in $t$-ZrO$_2$. The overall dielectric constants for
$c$-HfO$_2$ and $m$-HfO$_2$ are found to become smaller, while
$t$-HfO$_2$ has a much greater dielectric constant, than 
the corresponding values in ZrO$_2$. Moreover, our Raman results can 
be used in resolving the puzzles associated with the 
Raman spectrum for $m$-HfO$_2$.

This work has been supported by NSF Grant DMR-99-81193.
We wish to thank E. Garfunkel for useful discussions.


\begin{references}
\bibitem{explan-strain} All three HfO$_2$ phases are non-piezoelectric,
so the fixed-strain and free-stress dielectric tensors are
identical for these systems.
\end{references}
\end{document}